%% file: main.tex
\documentclass[conference]{IEEEtran}
\IEEEoverridecommandlockouts

\usepackage{cite}
\usepackage{amsmath,amssymb,amsfonts}
\usepackage{algorithm}
\usepackage{algorithmic}
\usepackage{CJKutf8}
\usepackage{graphicx}
\usepackage{array}
\usepackage{tabularx}
\usepackage{textcomp}
\usepackage{xcolor}
\usepackage{hyperref}
\def\BibTeX{{\rm B\kern-.05em{\sc i\kern-.025em b}\kern-.08em
    T\kern-.1667em\lower.7ex\hbox{E}\kern-.125emX}}
\makeatletter
\newcommand{\linebreakand}{%
  \end{@IEEEauthorhalign}
  \hfill\mbox{}\par
  \mbox{}\hfill\begin{@IEEEauthorhalign}
}
\makeatother
\begin{document}

\title{MAGE-RAG: Multigranular Adaptive Graph Evidence for Agentic Multimodal RAG in Long-Document QA}

\author{\IEEEauthorblockN{Yilong Zuo\textsuperscript{\textdagger}}
\IEEEauthorblockA{\textit{Beijing Institute of Technology}\\
Beijing, China \\
ORCID: 0009-0000-9350-1397}
\and
\IEEEauthorblockN{Xunkai Li\textsuperscript{\textdagger}}
\IEEEauthorblockA{\textit{Beijing Institute of Technology}\\
Beijing, China \\
ORCID: 0000-0002-1230-7603}
\and
\IEEEauthorblockN{Jing Yuan}
\IEEEauthorblockA{\textit{Beijing Institute of Technology}\\
Beijing, China \\
ORCID: 0009-0001-8379-4664}
\linebreakand
\IEEEauthorblockN{Qiangqiang Dai}
\IEEEauthorblockA{\textit{Beijing Institute of Technology}\\
Beijing, China \\
ORCID: 0000-0002-8569-6558}
\and
\IEEEauthorblockN{Hongchao Qin}
\IEEEauthorblockA{\textit{Beijing Institute of Technology}\\
Beijing, China \\
ORCID: 0000-0003-4364-0633}
\and
\IEEEauthorblockN{Ronghua Li}
\IEEEauthorblockA{\textit{Beijing Institute of Technology}\\
Beijing, China \\
ORCID: 0000-0001-8658-6599}
\thanks{\textsuperscript{\textdagger}Contributed equally to this work.}
}

\maketitle

\begin{CJK*}{UTF8}{gbsn}

\begin{abstract}
Long-document multimodal question answering requires a system to locate sparse evidence in long PDFs and integrate clues from text, tables, images, charts, and complex layouts. Existing RAG methods mostly rely on fixed Top-\(k\) retrieval over text chunks or pages. Text retrieval can compress the context but often loses visual and layout information; page-level visual retrieval preserves the original page, yet it also sends large irrelevant regions to the reader, leading to a static trade-off among evidence coverage, noise, and inference cost. This paper proposes MAGE-RAG, a multigranular adaptive graph evidence framework for long-document multimodal QA. MAGE-RAG uses page retrieval as the entry point for query-time evidence construction. Offline, it builds an evidence graph with page nodes and element nodes, encoding containment, reading order, layout adjacency, section hierarchy, and semantic-neighbor relations. At query time, an online evidence controller iteratively activates, opens, searches, and prunes evidence under explicit budgets. The resulting evidence subgraph is then rendered into structured multimodal reader input, allowing the LVLM to consume compact and relevant evidence within a limited context. On LongDocURL and MMLongBench-Doc, we establish a unified comparison and analysis protocol covering Direct MLLM, Text RAG, Page-level Visual RAG, and Graph/Agentic RAG. Experiments show that MAGE-RAG achieves 52.75 overall accuracy on LongDocURL, and 53.26 accuracy with 51.19 F1 on MMLongBench-Doc. Fine-grained breakdowns, budget-performance curves, ablations, and trace-based analysis further show that query-time evidence subgraph construction can balance dispersed evidence coverage with context-noise control. Our code is available at \href{https://github.com/laonuo2004/MAGE-RAG.git}{MAGE-RAG}.
\end{abstract}


\begin{figure}[!t]
    \centering
    \includegraphics[width=0.75\columnwidth]{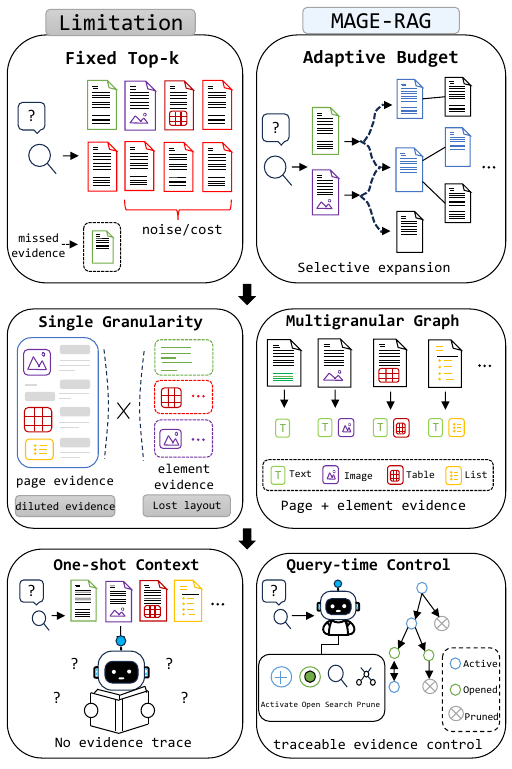}
    \caption{Motivating comparison between fixed-granularity retrieval and MAGE-RAG. Fixed Top-\(k\) retrieval can miss dispersed evidence while adding noisy pages, single-granularity evidence either dilutes local details or loses page layout, and one-shot contexts provide little control over the evidence construction process. MAGE-RAG instead uses an adaptive budget, a multigranular page--element graph, and query-time evidence control to selectively activate, open, search, and prune evidence.}
    \label{fig:intro_overview}
\end{figure}

\input{sections/introduction}

\input{sections/related_work}

\input{sections/problem_formulation}

\input{sections/empirical_study}

\newpage
\input{sections/methodology}

\input{sections/experiments}

\input{sections/conclusion}

\input{sections/acknowledgment}

\end{CJK*}

\bibliographystyle{IEEEtran}
\bibliography{references}

\end{document}

%% file: sections/introduction.tex
\section{Introduction}

Long-document multimodal question answering requires a system to answer natural-language questions over a PDF. Early benchmarks such as DocVQA, InfographicVQA, and ChartQA have shown that document QA must jointly handle text recognition, layout understanding, chart reasoning, and visual grounding~\cite{docvqa,infographicvqa,chartqa}. Document models such as LayoutLM and Donut further indicate that text, layout, and visual signals need to be modeled together~\cite{layoutlm,donut}. More recent benchmarks, including LongDocURL and MMLongBench-Doc, show that the task has moved beyond single-page understanding toward long-document analysis that is closer to realistic reading scenarios~\cite{longdocurl,mmlongbenchdoc}. Unlike short-document or text-only QA, a long PDF often contains tens or even hundreds of pages. Its evidence is sparse, and answer clues may be scattered across paragraphs, tables, charts, figures, section hierarchies, and page layouts. Some questions depend only on a local fact from a single page. Others require cross-page comparison, tracking context along document structure, or cross-checking visual content against extracted text. A system therefore has to find evidence that is sufficiently covering, compactly represented, and reliably usable by a multimodal reader, all under limited context and computation budgets.

Retrieval-augmented generation (RAG) mitigates the limitations of parametric memory by introducing external retrieval before generation~\cite{lewis2020rag,rag_survey}, and has been extended to visual document retrieval and multimodal document QA~\cite{colpali,visrag,m3docrag}. Existing multimodal RAG methods, however, are still largely organized around fixed retrieval units and one-shot context concatenation. This design leads to three related limitations. \textbf{L1: Fixed Top-$k$ and static budgets.} Fixed Top-$k$ retrieval maps every question to the same number of pages or chunks, making it difficult to handle single-page questions, cross-page questions, and unanswerable questions at the same time. Long-context RAG studies such as LongRAG also observe that retrieval granularity and context length jointly affect coverage and noise~\cite{longrag}. A small $k$ can miss dispersed evidence, whereas a large $k$ introduces redundant content and weakens the reader's attention. \textbf{L2: Single-granularity evidence.} Text or OCR chunk retrieval provides fine-grained local evidence, but may lose page-level visual cues, table structure, and layout relations. Page-level visual retrieval preserves the whole page, yet sends many irrelevant regions to the reader, causing evidence noise and input cost to grow quickly with Top-$k$~\cite{bm25,colbertv2,colpali}. \textbf{L3: Lack of query-time evidence control.} Existing pipelines usually pass retrieved candidates directly to the reader. They lack an explicit evidence state and trace that record which clues have been observed, which gaps still need to be searched, and which evidence should be expanded or pruned. Long-document multimodal QA therefore calls for a query-time evidence organization mechanism that can dynamically select, expand, preserve, and remove evidence under a budget constraint.

Graph structures, active retrieval, and agentic reading provide useful inspiration for organizing evidence in long contexts~\cite{react,flare,selfrag,graphreader,gretriever,graph_rag_survey,mldocrag,g2reader}. This work studies the following question: \textit{How can a system decide, for the current question, which evidence in a long document should be expanded, retained, or pruned?} To answer it, we formulate long-document multimodal QA as \textbf{query-time evidence management under a budget constraint}. The system maintains a traversable evidence structure over pages, regions, text blocks, tables, and image elements, and dynamically selects, expands, and materializes a compact evidence subgraph for the given query. Based on this formulation, MAGE-RAG (Multigranular Adaptive Graph Evidence for Retrieval-Augmented Generation) organizes long-document reading as a query-time evidence construction process. In the offline stage, it builds a document evidence graph that places full-page views and within-page local elements in the same evidence space. At query time, page-level visual retrieval provides the initial grounding pages, after which an online evidence controller iteratively performs evidence activation, opening, search, and pruning according to the current state. The resulting evidence subgraph is then rendered into a page-organized structured multimodal reader input.

MAGE-RAG separates the retrieval entry point, the query-time execution plan, and the final answer context into three connected components. Page-level retrieval provides initial grounding. The evidence graph supplies a document organization that can be expanded along structural and semantic relations. The controller updates the evidence state based on the current evidence state and recent trace, while the reader consumes the final materialized multimodal evidence. Through this process, coarse page context and fine-grained element evidence can be selected, pruned, and rendered within a shared state space.

\textbf{Contributions.} The contributions of this work are threefold. \textbf{(1) Problem formulation and motivation diagnosis.} We analyze the limitations of fixed Top-$k$, single-granularity retrieval, and the lack of query-time state management from the perspective of evidence organization, and formulate long-document multimodal QA as budgeted evidence subgraph construction. \textbf{(2) MAGE-RAG framework.} MAGE-RAG combines page-level visual entry points, a multigranular document graph, an online evidence controller, and structured reader rendering to select, prune, and materialize page context and element-level local evidence within the same stateful process. \textbf{(3) Unified experimental protocol and auditable analysis.} We build a unified reader protocol around LongDocURL and MMLongBench-Doc, and analyze the benefits and failure modes of different evidence organization strategies in terms of overall accuracy, evidence budget, single-page and cross-page questions, evidence sources, and trace behavior.

%% file: sections/related_work.tex
\section{Related Work}

\textbf{Long-document multimodal question answering.}
Document understanding is moving from single-page forms, receipts, and webpage screenshots toward long-PDF reading. Early visual document QA benchmarks such as DocVQA, InfographicVQA, and ChartQA advanced the field from the perspectives of scanned documents, infographics, and chart question answering~\cite{docvqa,infographicvqa,chartqa}. Models such as LayoutLM and Donut further study page content through text-layout pre-training and OCR-free generative document understanding~\cite{layoutlm,donut}. Recent benchmarks make the long-document setting more explicit. MMLongBench-Doc annotates dimensions such as evidence source, evidence pages, cross-page questions, and unanswerable questions~\cite{mmlongbenchdoc}; LongDocURL covers document understanding, numerical reasoning, and cross-element localization, emphasizing the need to locate and integrate dispersed evidence in large multimodal documents~\cite{longdocurl}. These settings shift evaluation from single-page content recognition to evidence localization, cross-page integration, and multimodal information alignment over a long document. Beyond the final answer, single-page/cross-page behavior, evidence source, answer format, and evidence budget have also become necessary dimensions for analyzing long-document multimodal QA.

\textbf{From text retrieval to visual document RAG.}
A common document RAG pipeline first parses PDF pages into OCR text, and then uses sparse or late-interaction retrievers such as BM25 and ColBERTv2 to retrieve text chunks~\cite{bm25,colbertv2}. Such text-only pipelines produce compact textual contexts, but page layout, table structure, chart geometry, and text-visual alignment can be weakened during parsing. General RAG studies also show that retrieval granularity, context length, and dynamic retrieval all affect generation quality~\cite{rag_survey,longrag,flare,selfrag}. To preserve visual and layout cues from the original page, recent work has moved further toward page-level visual retrieval. ColPali encodes pages with multi-vector visual representations and computes query-page relevance~\cite{colpali}. VisRAG, M3DocRAG, VDocRAG, and EVisRAG extend visual document RAG from the perspectives of multimodal retrieval, multi-page DocVQA, and multi-image reasoning~\cite{visrag,m3docrag,vdocrag,evisrag}. Meanwhile, AgenticOCR and RegionRAG show that static parsing after whole-page retrieval can still pass many irrelevant regions to the generator, making region-level or on-demand parsing important~\cite{agenticocr,regionrag}. MAGE-RAG preserves the stable entry point of page retrieval, but maintains a multigranular evidence graph over the retrieved pages. The query-time controller then decides which local evidence should be opened, retained, or pruned.

\textbf{Structure-aware and agentic evidence organization.}
Another line of work treats long-context QA as a structured exploration problem. ReAct presents a general paradigm in which language models alternate between reasoning and acting~\cite{react}. FLARE and Self-RAG further suggest, from active retrieval and self-reflection perspectives, that retrieval need not be limited to a single upfront step~\cite{flare,selfrag}. GraphReader organizes long text as a graph and lets an agent progressively visit nodes and their neighbors~\cite{graphreader}. G-Retriever formulates graph retrieval as subgraph selection for textual graph QA~\cite{gretriever}, and the Graph RAG survey summarizes how graph structures support knowledge organization in retrieval-augmented generation~\cite{graph_rag_survey}. For visually rich documents, MLDocRAG organizes paragraphs, images, and tables with query-centric semantic nodes~\cite{mldocrag}; LAD-RAG combines a layout-aware symbolic document graph with neural and symbolic indexes~\cite{ladrag}; and $G^2$-Reader maintains both a content graph and a planning graph for multi-step evidence completion~\cite{g2reader}. More recently, FlowReader formulates evidence assembly in multimodal long-document QA as a minimum-cost flow problem~\cite{flowreader}. These studies move long-document QA from one-shot retrieval toward structured and stateful evidence construction. MAGE-RAG uses the graph as the query-time processing space and defines the final context as a materialized evidence subgraph under budget constraints. The controller updates this subgraph, rather than directly taking on free-form answering.

%% file: sections/problem_formulation.tex
\section{Problem Formulation}

This work studies long-document multimodal QA within a given document. Given a PDF document \(D\) and a natural-language question \(q\), the system is expected to generate an answer \(y\) using only the evidence visible in \(D\). Unlike conventional text QA, \(D\) often contains many pages, and the relevant clues may appear in page screenshots, paragraph text, tables, figures, images, heading structures, reading order, and page layout. This setting inherits the central requirement of visual document QA: text, layout, and visual evidence must be modeled jointly~\cite{docvqa,layoutlm,donut}. Since the full document usually cannot be passed directly to the reader, the system needs to progressively select, expand, prune, and finally materialize a query-specific evidence subgraph under a limited budget. We formalize this process as query-time evidence subgraph construction under budget constraints.

\textbf{Multigranular Document Evidence Graph.}
Let \(D=\{P_i\}_{i=1}^{n}\) denote a document with \(n\) pages. After offline parsing, the document is represented as a heterogeneous evidence graph:
\begin{equation}
    \mathcal{G}_D=(\mathcal{V}, \mathcal{E}),
    \label{eq:evidence_graph}
\end{equation}
where \(\mathcal{V}=\mathcal{V}^{p}\cup\mathcal{V}^{e}\). Here, \(\mathcal{V}^{p}=\{v_i^p\}_{i=1}^{n}\) denotes page nodes, while \(\mathcal{V}^{e}\) denotes in-page element nodes, such as headings, paragraphs, lists, tables, images, charts, formulas, code blocks, and algorithm blocks. Each node \(v\in \mathcal{V}\) is associated with a multimodal description
\(\mathbf{x}_v=(a_v,t_v,b_v,m_v)\), where \(a_v\) is the node summary, \(t_v\) is the available text or structured content, \(b_v\) is the bounding box within the page, and \(m_v\) is a page image or local image reference. Not every field is required. For example, a page node usually keeps the full-page image and page summary, whereas a table or chart node may contain text, a bbox, and a cropped screenshot at the same time.

The edge set \(\mathcal{E}\subseteq \mathcal{V}\times\mathcal{R}\times\mathcal{V}\) encodes structural relations inside the document, where \(\mathcal{R}\) is the set of relation types, including containment, reading order, layout, section hierarchy, and semantic relations. The graph serves as the search space for organizing evidence: page nodes provide stable visual grounding, element nodes provide denser local evidence, and edges describe the document structure that can be traversed during query-time expansion.

\textbf{Query-Time Evidence State.}
For a question \(q\), the system maintains an evidence state over \(\mathcal{G}_D\) that evolves across iterations. Let
\begin{equation}
    s_t(v)\in \{\mathsf{Inactive}, \mathsf{Active}, \mathsf{Opened}, \mathsf{Pruned}\}
    \label{eq:node_state}
\end{equation}
denote the state of node \(v\) after iteration \(t\). \(\mathsf{Inactive}\) means that the node has not entered the current working memory. \(\mathsf{Active}\) means that the node has been selected into the working memory and participates in later decisions and final rendering through its summary, type, and preview. \(\mathsf{Opened}\) means that, beyond its summary-level representation, the node exposes detailed content, bbox information, and available images. \(\mathsf{Pruned}\) means that the node is judged low-value or distracting for the current question. The working evidence subgraph induced by active and opened nodes is denoted as
\begin{equation}
\begin{aligned}
    \mathcal{V}_{t,\mathrm{work}}^q=\{v\mid s_t(v)\in\{\mathsf{Active},\mathsf{Opened}\}\},
    \\
    \mathcal{G}_{t,\mathrm{work}}^q=\mathcal{G}_D[\mathcal{V}_{t,\mathrm{work}}^q].
\end{aligned}
    \label{eq:query_subgraph}
\end{equation}
Here, \(\mathcal{G}_{t,\mathrm{work}}^q\) is the working evidence subgraph constructed so far for the current question. The reader does not consume the bare subgraph. Instead, it receives a multimodal input rendered from the final state, written as \(X_{\mathrm{read}}=\operatorname{Render}(q,\mathcal{G}_D,S_T)\). This state-based definition separates clues that have been located but not fully expanded from evidence that has been opened and can support answer verification. Pruning is also recorded explicitly, rather than being treated as a passive loss of pages or blocks during context concatenation.

\textbf{Action Space and Budget.}
Query-time evidence construction is carried out through a sequence of actions. This work considers the following action space:
\begin{equation}
    \begin{aligned}
    \mathcal{A}=\{&
    \textsc{ActivatePage}, \textsc{ActivateNode}, \textsc{OpenNode},\\
    &\textsc{SearchEvidence}, \textsc{PruneNode}, \textsc{Stop}\}.
    \end{aligned}
    \label{eq:action_space}
\end{equation}
\textsc{ActivatePage} adds a page node to the working memory as an entry point. \textsc{ActivateNode} activates candidate elements along containment, layout, reading order, section hierarchy, or semantic relations. \textsc{OpenNode} expands an activated non-page node. \textsc{SearchEvidence} performs focused re-retrieval within the same document according to the current evidence gap. \textsc{PruneNode} marks a low-value node as pruned, and \textsc{Stop} terminates expansion and moves the process to the reader stage. All actions are executed within the allowed page range of the current document and introduce no external knowledge.

Because the reader has limited context length, image capacity, and model-call budget, the action sequence must satisfy budget constraints. We use
\[
    B=(B_p,B_a,B_A,B_o,B_i,B_T,B_c)
\]
to denote the budget vector, where \(B_p\) controls the number of initial page entries, \(B_a\) controls the number of numbered activation actions allowed in one iteration, \(B_A\) controls the total number of numbered activation actions, \(B_o\) controls the number of nodes that can be finally opened, \(B_i\) controls the number of images or content parts accepted by the reader, \(B_T\) controls the number of controller iterations, and \(B_c\) upper-bounds the total input cost or latency cost. A practical system may instantiate only part of this vector. In the implementation of MAGE-RAG, the hard constraints mainly apply to page entries, numbered activation actions, iteration count, and final-open nodes; reader input size and total cost are recorded and analyzed as evaluation statistics. These constraints prevent the evidence size from growing without bound as Top-\(k\) or document length increases.

\textbf{Objective.}
Ideally, the system should construct reader evidence that is both answer-supporting and compact under budget \(B\). The following objective characterizes the problem, rather than a trainable loss directly optimized in this work:
\begin{equation}
\begin{aligned}
    \max_{\pi} \quad &
    \mathrm{Ans}(q, X_{\mathrm{read}})
    - \lambda\,\mathrm{Noise}(X_{\mathrm{read}})
    - \mu\,\mathrm{Cost}(X_{\mathrm{read}}) \\
    \mathrm{s.t.}\quad &
    S_T=\mathrm{Apply}(\mathcal{G}_D,q,\pi),\\
    &
    X_{\mathrm{read}}=\operatorname{Render}(q,\mathcal{G}_D,S_T),\\
    &
    \mathrm{Budget}(\pi,S_T,X_{\mathrm{read}})\le B .
\end{aligned}
\label{eq:objective}
\end{equation}
Here, \(\pi\) is the query-time policy, \(S_T\) is the final evidence state induced by the action sequence, \(\mathrm{Ans}(\cdot)\) measures how well the rendered evidence supports the answer, \(\mathrm{Noise}(\cdot)\) measures irrelevant or distracting evidence, and \(\mathrm{Cost}(\cdot)\) captures the inference cost introduced by pages, nodes, images, and tokens. The coefficients \(\lambda\) and \(\mu\) control the importance of evidence purity and cost. Eq.~\eqref{eq:objective} is used to define the target of the problem. In practice, answerability and evidence noise usually lack node-level annotations, and the required amount of evidence varies substantially across questions. This work therefore does not assume that an end-to-end optimizer can be directly trained for the objective.

\textbf{Solving Strategy.}
MAGE-RAG approximates Eq.~\eqref{eq:objective} as an online control process. Page-level visual retrieval first provides \(\pi\) with initial entry pages, avoiding a difficult search from scratch in a long document. Then, the candidate generator produces feasible actions from the current state and graph relations. The evaluator estimates the marginal utility of each action from the question, current evidence state, and recent trace, and executes activation, opening, search, or pruning under the budget constraints. Finally, non-pruned page and element nodes are rendered into structured multimodal reader evidence, from which the reader generates the answer. Under this formulation, retrieval provides entry points, the graph structure provides the expansion space, the state machine records query-time decisions, the budget controls evidence scale, and the reader consumes the final rendered \(X_{\mathrm{read}}\).

%% file: sections/empirical_study.tex
\section{Empirical Study}

This section presents a design-oriented empirical study on two limitations in long-document multimodal QA: fixed Top-\(k\) retrieval and single evidence granularity. The goal is not to build another leaderboard. Instead, the analysis uses controlled comparisons to expose how different evidence organization strategies behave, and connects these observations to the three design choices in Eq.~\eqref{eq:objective}: a page-level entry, a multigranular evidence graph, and a query-time evidence controller.

\textbf{Research Questions.}
The analysis is organized around three questions. \textbf{Q1:} Can fixed Top-\(k\) page retrieval serve both single-page and cross-page questions? \textbf{Q2:} Is a single-granularity context sufficient for heterogeneous evidence such as text, tables, charts, figures, images, and layout cues? \textbf{Q3:} Can query-time expansion, opening, search, and pruning provide an interpretable evidence organization process, rather than merely increasing the input size? These questions correspond to the sparsity, cross-page dependency, and multimodality emphasized by recent long-document multimodal benchmarks~\cite{mmlongbenchdoc,longdocurl}, and are also related to evidence organization in graph-based and agentic document reading methods~\cite{mldocrag,g2reader}.

\textbf{Datasets and Protocol.}
The empirical study focuses on LongDocURL and MMLongBench-Doc. Table~\ref{tab:benchmark_protocol} summarizes their basic statistics and evaluation protocols. The main experiments later use a unified reader model, retrieval scope, and answer evaluation protocol. This section instead relies on controlled MAGE-RAG runs and stratified observations to motivate the page-level entry, multigranular evidence, and query-time control.

\begin{table*}[t]
\centering
\caption{Benchmark statistics and evaluation protocols used in the empirical study and experiments. Evidence-page statistics for LongDocURL are derived from sample annotation fields; in MMLongBench-Doc, samples with empty evidence pages correspond to unanswerable questions.}
\label{tab:benchmark_protocol}
\footnotesize
\renewcommand{\arraystretch}{1.18}
\begin{tabularx}{\textwidth}{@{}lcc>{\raggedright\arraybackslash}X>{\raggedright\arraybackslash}X>{\raggedright\arraybackslash}X>{\raggedright\arraybackslash}X@{}}
\hline
Benchmark & Docs & QA & Groups & Evidence pages & Evidence types & Metrics \\
\hline
LongDocURL & 396 & 2,325 &
Understanding 1,243; locating 695; reasoning 387 &
1 page 1,093; \(\geq\)2 pages 1,230; missing 2 &
Text, table, figure, layout, and mixed formats &
Overall acc.; task, element, and page breakdowns \\
MMLongBench-Doc & 135 & 1,091 &
Research report, academic paper, guidebook, tutorial, financial report &
Single 485; cross 360; unanswerable 246 &
Pure text, layout text, table, chart, figure, and mixed formats &
Overall acc./F1; single, cross, and unanswerable breakdowns \\
\hline
\end{tabularx}
\end{table*}

\textbf{Fixed Top-\(k\) makes coverage and noise difficult to control jointly.}
\input{tables/mmlongbench_magerag_topk_table}

Table~\ref{tab:magerag_topk} fixes the controller, action budget, and graph structure on MMLongBench-Doc, while varying only the initial page-level entry \(k\). Increasing Top-\(k\) can recover some cross-page evidence, but it is not a reliably monotonic solution. From \(k=1\) to \(k=5\), both overall and cross-page metrics improve, whereas the cross-page metric drops at \(k=8\). A larger value, \(k=10\), further improves F1 and the cross-page metric, but overall accuracy decreases and the unanswerable metric continues to fall. Fixed Top-\(k\) visual RAG therefore faces a trade-off across question types: a small \(k\) may miss dispersed evidence, while a large \(k\) can disturb the calibration between answerable and unanswerable samples and pass more low-density pages to the reader. A better treatment is to keep an extensible page-level entry, while letting the query-time evidence state decide the final context.

\textbf{Stratified results reveal differences in evidence granularity.}
The stratified metrics of LongDocURL offer another perspective, as they distinguish both task types and evidence element types. In MAGE-RAG runs, single-page questions still outperform multi-page questions, and locating remains lower than understanding and reasoning. One plausible explanation is that text paragraphs often contain the answer string directly, tables and charts require local visual verification, and locating questions additionally depend on titles, layout, and page-level context. Page-level evidence preserves the full visual context, but it also asks the reader to localize the relevant region again within a whole-page image. Text or chunk evidence is more compact, but may lose layout, chart structure, and visual alignment. MAGE-RAG therefore keeps both page nodes and element nodes: the former provide visual anchors, the latter expose high-density local evidence such as tables, charts, titles, and paragraphs, and the evidence graph connects the two granularities at query time.

This granularity gap is also reflected in the stratified results of the main experiments. Compared with uniformly reproduced text RAG and page-level visual RAG, MAGE-RAG uses page and element evidence more stably, although cross-page evidence organization remains a major bottleneck. The observation here is not that single-granularity methods must fail in every subgroup. Rather, long-document multimodal QA requires both page-level visual anchors and element-level local evidence.

\textbf{Query-time expansion reflects evidence-scale differences across questions.}
Different questions require different amounts of evidence, so a single static number of pages cannot fully explain system behavior. The controller produces an auditable trace, including activated pages, opened nodes, search queries, pruned nodes, and stopping reasons. For failed samples, these traces can help case analysis and system debugging by indicating whether the bottleneck comes from missed initial pages, insufficient graph node quality, premature stopping, repeated search, or reader failure over the final evidence. This section first reports behavior statistics at the aggregate level.

\begin{figure}[t]
\centering
\includegraphics[width=\columnwidth]{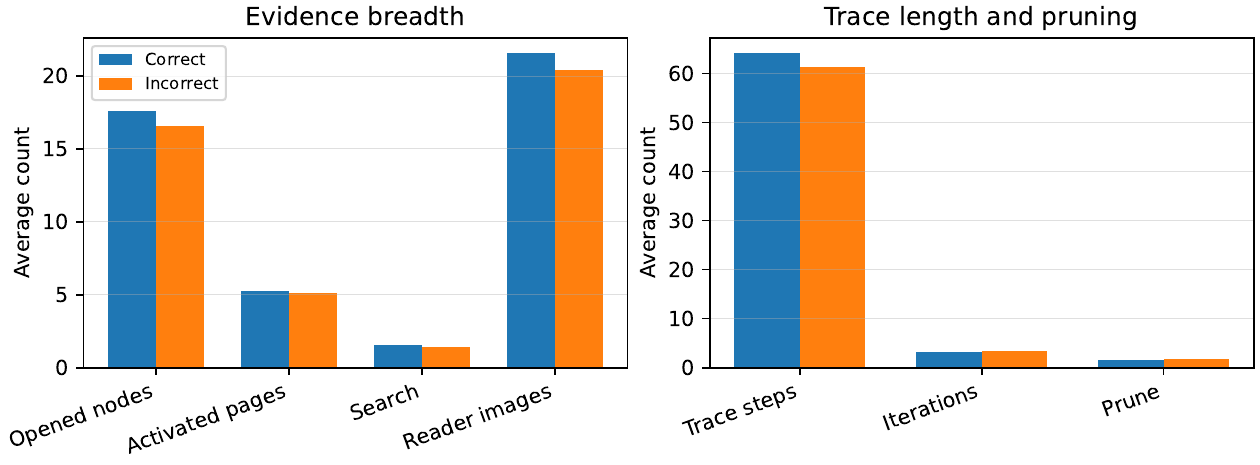}
\caption{Trace behavior statistics from a representative MAGE-RAG run on MMLongBench-Doc. The left panel compares the evidence width of correct/incorrect samples, while the right panel compares trace length, iteration count, and pruning actions.}
\label{fig:magerag_trace_stats}
\end{figure}

Figure~\ref{fig:magerag_trace_stats} shows that correct samples do not gain their advantage from substantially longer traces. Instead, under a similar iteration budget, they open slightly more evidence nodes, activate slightly more pages, and trigger slightly more search actions. Incorrect samples have marginally higher average iteration counts and pruning counts. Their failures do not necessarily come from insufficient exploration; they may also result from low-quality candidate evidence, missed search results, or a controller that still fails to form an effective evidence state along a longer trajectory. Such statistics turn the evidence construction behavior of correct/incorrect samples into diagnostic signals.

\textbf{Implications for MAGE-RAG.}
The above observations motivate the three designs of MAGE-RAG. Page-level visual retrieval serves as the entry point and provides relatively robust initial grounding for long PDFs. The retrieval results initialize a multigranular evidence state, allowing the system to expose page evidence and element evidence at different levels of detail. The explicit controller and budget constraints then handle evidence-scale differences across questions. In this view, retrieval is not a one-shot decision about how many pages to pass to the reader, but the initialization of a working evidence state that can be expanded, opened, searched, and pruned according to the current question. The next section presents the graph construction, state update, and reader rendering designs of MAGE-RAG.

%% file: tables/mmlongbench_magerag_topk_table.tex
\begin{table}[t]
\centering
\caption{Performance with different initial page Top-\(k\) values on MMLongBench-Doc. Except for Top-\(k\), all other budgets and graph settings are kept fixed; Acc/F1 are reported as percentages.}
\label{tab:magerag_topk}
\scriptsize
\resizebox{\columnwidth}{!}{%
\begin{tabular}{cccccc}
\hline
Top-\(k\) & Acc & F1 & Single & Cross & Unans. \\
\hline
1 & 48.95 & 45.75 & 55.87 & 24.42 & 73.77 \\
3 & 52.25 & 49.44 & 59.43 & 31.10 & 70.90 \\
5 & 53.05 & 51.12 & 60.16 & 34.42 & 67.62 \\
8 & 51.53 & 49.34 & 61.41 & 30.36 & 65.16 \\
10 & 50.10 & 58.61 & 57.97 & 37.55 & 62.29 \\
\hline
\end{tabular}%
}
\end{table}

%% file: sections/methodology.tex
\section{Methodology}
\label{sec:methodology}

\begin{figure*}[t]
    \centering
    \includegraphics[width=0.998\textwidth]{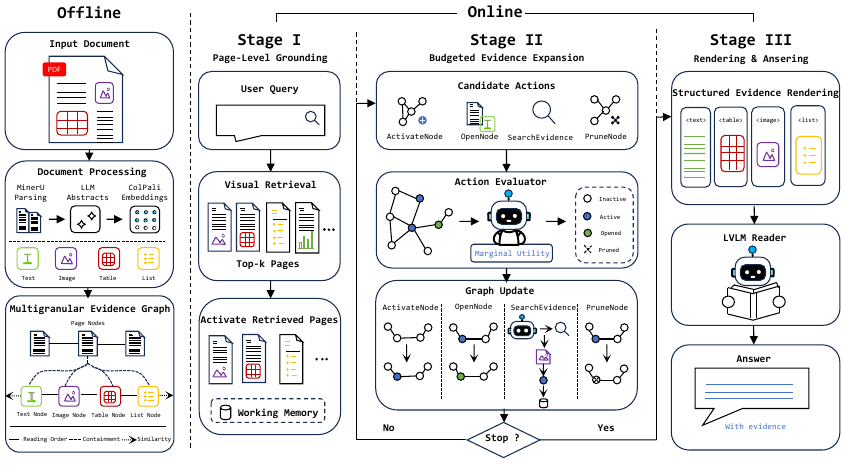}
    \caption{\textbf{Overall framework of MAGE-RAG.} The system first constructs an offline multigranular evidence graph with page nodes and element nodes. At query time, page-level visual retrieval provides entry pages, after which the online evidence controller activates, opens, searches, and prunes evidence on the graph. The resulting evidence subgraph is finally rendered into structured multimodal reader input.}
    \label{fig:magerag_framework}
\end{figure*}

This section presents the design of MAGE-RAG. Fig.~\ref{fig:magerag_framework} illustrates the overall workflow. The framework contains four components: offline multigranular evidence graph construction, page-level initial grounding, an online evidence controller, and structured multimodal reader rendering. The first three components determine which evidence enters the working memory, while the last component determines how this evidence is consumed by the LVLM reader. Algorithm~\ref{alg:magerag} summarizes the query-time evidence construction process.

\begin{algorithm}[H]
\caption{MAGE-RAG Query-Time Evidence Construction}
\label{alg:magerag}
\small
\begin{algorithmic}[1]
\REQUIRE Question \(q\), document \(D\), graph \(\mathcal{G}_D\), budgets \(k,a,A_{\max},B_T,B_o\)
\ENSURE Answer \(y\) and evidence trace \(\tau\)
\STATE \(\mathcal{P}_0 \leftarrow \operatorname{TopKPageGrounding}(q,D,k)\)
\STATE \(S \leftarrow \operatorname{ActivatePages}(\mathcal{P}_0)\), \(\tau \leftarrow [\,]\)
\FOR{\(t=1\) to \(B_T\)}
    \STATE \(\mathcal{C} \leftarrow \operatorname{GenerateCandidates}(\mathcal{G}_D,S)\)
    \STATE \(A \leftarrow \operatorname{Eval}_{\theta}(q,S,\mathcal{C},\tau)\)
    \STATE \(A \leftarrow \operatorname{ApplyActivationBudget}(A,a,A_{\max})\)
    \IF{\(A\) only contains \textsc{Stop}}
        \STATE \(\tau \leftarrow \operatorname{UpdateTrace}(\tau,A,S)\)
        \STATE \textbf{break}
    \ENDIF
    \STATE \(S \leftarrow \operatorname{ApplyActions}(S,A)\)
    \STATE \(\tau \leftarrow \operatorname{UpdateTrace}(\tau,A,S)\)
\ENDFOR
\STATE \(S^\star \leftarrow \operatorname{FinalOpenActiveEvidence}(S,B_o)\)
\STATE \(X_{\mathrm{read}} \leftarrow \operatorname{Render}(q,\mathcal{G}_D,S^\star)\)
\STATE \(y \leftarrow \operatorname{Reader}(X_{\mathrm{read}})\)
\RETURN \(y,\tau\)
\end{algorithmic}
\end{algorithm}

\subsection{Offline Multigranular Evidence Graph Construction}
\label{subsec:offline_graph}

\textbf{Node construction.}
Given a PDF document \(D\), the offline pipeline first uses a document parser to obtain page screenshots, structured in-page blocks, bounding boxes, captions, footnotes, table HTML, and available text. It then constructs two types of nodes. A page node \(v_i^p\) represents the \(i\)-th page and stores the page summary, page-image path, and page-level embedding path. An element node \(v_j^e\) represents fine-grained evidence inside a page, including headings, paragraphs, lists, tables, images, charts, formulas, code blocks, and algorithm blocks. Each element node stores its parent page, local bounding box, type field, available text or structured content, and an optional local-image path.

This node design preserves two evidence granularities. Page nodes provide stable retrieval units, since a long PDF usually has far fewer pages than elements and full-page images preserve complete layout and visual context. Element nodes provide denser local evidence, allowing the controller to open only the paragraphs, tables, or charts that are relevant to the question. The offline stage also generates short abstracts for pages and elements, which serve as compressed semantic representations during online decision making.

\textbf{Relation construction.}
Over the node set \(\mathcal{V}^{p}\cup\mathcal{V}^{e}\), MAGE-RAG constructs several types of intra-document relations:
\begin{equation}
    \mathcal{E}
    =
    \mathcal{E}_{\mathrm{con}}
    \cup \mathcal{E}_{\mathrm{read}}
    \cup \mathcal{E}_{\mathrm{lay}}
    \cup \mathcal{E}_{\mathrm{sec}}
    \cup \mathcal{E}_{\mathrm{sem}} .
    \label{eq:method_edge_union}
\end{equation}
Here, \(\mathcal{E}_{\mathrm{con}}\) denotes page-to-element containment relations; \(\mathcal{E}_{\mathrm{read}}\) denotes reading-order relations between pages or between elements; \(\mathcal{E}_{\mathrm{lay}}\) captures layout relations such as left--right adjacency according to bounding boxes; \(\mathcal{E}_{\mathrm{sec}}\) connects section headings to their subheadings or content blocks according to the heading hierarchy; and \(\mathcal{E}_{\mathrm{sem}}\) builds semantic-neighbor relations from node embeddings. These edge types correspond to different reading behaviors. Containment supports moving from a page to local evidence, reading order supports sequential completion, layout relations support spatial alignment among charts, titles, and captions, section hierarchy supports expansion along document organization, and semantic relations provide cross-location semantic jumps.

\subsection{Page-Level Initial Grounding}
\label{subsec:initial_grounding}

Directly searching element nodes over the full graph in a long document leads to a large candidate space and substantial parsing noise. MAGE-RAG therefore uses page-level visual retrieval as the entry point for the online stage. Given a question \(q\), the system computes the relevance between the query and page embeddings within the allowed page set of the current document, and selects the Top-\(k\) pages:
\begin{equation}
    \mathcal{P}_0
    =
    \operatorname{TopK}_{P_i \in D}
    \; \mathrm{sim}(\mathbf{e}_q, \mathbf{e}_{P_i}; k).
    \label{eq:initial_pages}
\end{equation}
where \(\mathbf{e}_q\) and \(\mathbf{e}_{P_i}\) denote the vision-language retrieval representations of the query and page, respectively. The retrieved pages are added to the evidence state through \textsc{ActivatePage} and form the working memory at iteration 0. The controller then decides, based on these page entries, whether to open in-page elements, expand along graph relations, or launch a new focused search.

Compared with chunk retrieval, a page-level entry is coarser, but it is often more robust when parsing noise is strong or when the query wording does not match extracted text. Subsequent graph expansion moves from the entry pages to local nodes, compensating for the coarse granularity of page-level grounding.

\subsection{Online Evidence Controller}
\label{subsec:online_controller}

\textbf{Evidence state.}
The online stage maintains a mutable state \(S_t=\{s_t(v)\mid v\in\mathcal{V}\}\). Each node is initialized as \(\mathsf{Inactive}\). When a page or element is selected into the working memory, its state becomes \(\mathsf{Active}\). An active node enters the evaluator as lightweight evidence through its summary, type, preview, and available location information, and it can also provide summary-level context in the final reader input. When the detailed content, bounding box, or image reference of an element needs to be explicitly materialized, its state becomes \(\mathsf{Opened}\). Beyond summary-level context, an opened node exposes fuller text, structured content, and image or crop references. When the controller judges that a node consumes context or distracts from answering the question, its state becomes \(\mathsf{Pruned}\). Many pieces of evidence therefore remain as clues in early iterations and are opened only when their marginal utility is sufficiently high.

\textbf{Candidate generation.}
At iteration \(t\), the candidate generator enumerates feasible candidates from \(\mathcal{G}_D\) and \(S_{t-1}\). The candidates mainly come from three sources:
\begin{equation}
    \mathcal{C}_t
    =
    \Gamma_{\mathrm{page}}(S_{t-1})
    \cup
    \Gamma_{\mathrm{rel}}(S_{t-1},\mathcal{G}_D)
    \cup
    \Gamma_{\mathrm{search}}(S_{t-1}).
    \label{eq:candidate_union}
\end{equation}
\(\Gamma_{\mathrm{page}}\) contains inactive elements on activated pages, usually corresponding to containment expansion from a page to paragraphs, headings, tables, or charts. \(\Gamma_{\mathrm{rel}}\) contains neighboring nodes reached from activated elements through reading-order, layout, section-hierarchy, or semantic edges. \(\Gamma_{\mathrm{search}}\) denotes new entry candidates brought by the focused search in the previous iteration. In the current implementation, a search request first triggers page re-retrieval within the same document and activates newly retrieved pages; the next candidate-generation round then exposes elements on those pages. The candidate generator only enumerates possible actions. Whether the current evidence is sufficient is left to the evaluator.

\textbf{Marginal-utility decision.}
The evaluator receives the question, current evidence state, recent trace, and candidate-action list, and outputs a structured decision:
\begin{equation}
    A_t
    =
    \operatorname{Eval}_{\theta}
    (q, S_{t-1}, \mathcal{C}_t, \tau_{t-1}),
    \label{eq:evaluator_decision}
\end{equation}
where \(\tau_{t-1}\) is the recent action trace. In the implementation, the evaluator observes the page-organized evidence state through an XML prompt and can select existing \textsc{ActivateNode} candidates only by their indices. \textsc{OpenNode}, \textsc{SearchEvidence}, and \textsc{PruneNode} are instead requested through dedicated XML fields. This indexed interaction reduces errors caused by copying long node IDs or edge IDs, and it also makes every state transition recoverable from the trace.

The executable actions include five main operations. \textsc{ActivatePage} adds a page to the working memory as an entry point. \textsc{ActivateNode} marks a candidate element or relational neighbor as active. \textsc{OpenNode} expands an active non-page node. \textsc{SearchEvidence} launches focused re-retrieval within the allowed pages of the same document according to the current evidence gap, and uses the retrieved pages as new entries. \textsc{PruneNode} marks low-value pages or elements as pruned. The \textsc{Stop} action in Eq.~\eqref{eq:action_space} corresponds to the termination flag in the evaluator decision. Online expansion ends only when the evaluator returns stop and makes no activation, opening, search, or pruning request.

\textbf{Budget control.}
MAGE-RAG uses budget constraints to prevent the evidence state from growing without bound. Let \(k\equiv B_p\) be the number of initial page entries, \(a\equiv B_a\) the maximum number of indexed \textsc{ActivateNode} actions executed in one iteration, \(A_{\max}\equiv B_A\) the total upper bound on indexed activation actions, \(B_T\) the maximum number of controller iterations, and \(B_o\) the final-open node budget. The online construction process for one query can be summarized as:
\begin{equation}
\begin{aligned}
    S_t &= \operatorname{Apply}(S_{t-1}, A_t), \quad t=1,\ldots,B_T,\\
    |A_t^{\mathrm{act}}| &\le B_a,\quad
    \sum_{t}|A_t^{\mathrm{act}}| \le B_A,\quad
    |\mathcal{V}_{\mathrm{opened}}| \le B_o .
\end{aligned}
\label{eq:budget_constraints_method}
\end{equation}
Here, \(B_a\) and \(B_A\) constrain the \textsc{ActivateNode} actions selected by the evaluator through candidate indices. \textsc{OpenNode}, \textsc{SearchEvidence}, and \textsc{PruneNode} are triggered through dedicated fields and are controlled indirectly by the watchdog, the single-search-per-iteration rule, and the final rendering size. A repeated no-op watchdog handles consecutive invalid actions. After the loop ends, the system performs final opening for highly relevant active non-page nodes according to question relevance, adding clues that have been located but not yet opened to the final reader evidence.

\subsection{Structured Multimodal Reader Rendering}
\label{subsec:reader_rendering}

The online stage produces a stateful evidence subgraph, whereas the reader needs messages that can be directly fed into the LVLM. MAGE-RAG introduces a structured renderer that organizes non-pruned pages and elements into XML according to the page hierarchy, and aligns them with page images, local node images, or bounding-box crops. After final opening, the reader input is written as:
\begin{equation}
    X_{\mathrm{read}}
    =
    \operatorname{Render}
    (q, \mathcal{G}_D, S^\star)
    =
    [X_{\mathrm{xml}};\mathcal{I}_{\mathrm{page}};\mathcal{I}_{\mathrm{node}}].
    \label{eq:reader_render}
\end{equation}
where \(X_{\mathrm{xml}}\) contains the question, activated page indices, evidence-use policy, page abstracts, node types, node abstracts, and available bounding boxes; \(\mathcal{I}_{\mathrm{page}}\) denotes page screenshots; and \(\mathcal{I}_{\mathrm{node}}\) denotes local images of opened nodes or crops obtained from bounding boxes.

The renderer moves evidence organization into the construction of reader input. The reader receives an evidence context aligned by page, node, and image reference. For table, chart, image, and figure nodes, local images or crops are used to verify visual details. For paragraph and heading nodes, abstracts compress the textual context. When extracted text conflicts with visible image content, the reader is instructed to follow the visible document content. The final answer is generated only from the provided evidence; if the evidence is insufficient, the reader outputs \texttt{Not answerable.}.

\subsection{Analysis and Comparison}
\label{subsec:method_analysis}

\textbf{Complexity analysis.}
Suppose the offline evidence graph contains \(|\mathcal{V}|\) nodes and \(|\mathcal{E}|\) edges. Offline structural edges mainly come from page containment, reading order, layout adjacency, and section hierarchy. Semantic edges usually retain \(K_s\) nearest neighbors for each node, so the graph size is approximately \(\mathcal{O}(|\mathcal{V}|K_s+|\mathcal{E}_{\mathrm{struct}}|)\). The online stage still performs initial page retrieval within the allowed pages, but controller expansion and the final reader input do not traverse the full graph. Instead, local candidates are generated from currently active pages or active elements. If the number of candidates in each iteration is denoted by \(C_t\), the main invocation scale of the controller is related to \(\sum_{t=1}^{T} C_t\) and the number of rendered content parts. The budgets \(B_a,B_A,B_T,B_o\) restrict long-document reading to a finite-step evidence selection process.

\textbf{Comparison with fixed Top-\(k\) RAG.}
Fixed Top-\(k\) page RAG passes retrieval results directly to the reader, and its context size is determined by a global \(k\). MAGE-RAG uses \(k\) as an entry budget; the evidence that actually enters the reader is decided by the subsequent state machine. For single-page questions, the controller can stop after visiting only a few pages and nodes. For cross-page or cross-element questions, it can increase evidence coverage through relational expansion and focused search. The trade-off between coverage and noise is therefore moved into a per-question evidence-state update process.

\textbf{Comparison with single-granularity retrieval.}
Text/chunk RAG is compact, but it can lose layout information, chart structure, and visual alignment. Page-level visual RAG preserves visual context, yet it transfers the localization burden to the reader. The multigranular graph in MAGE-RAG places both granularities in the same evidence space: page nodes provide visual grounding, element nodes provide dense local evidence, and edges describe expandable relations between pages and elements. Depending on the question, the controller can switch expansion paths among page entries, local elements, and cross-location neighbors.

\textbf{Auditability.}
Actions such as page activation, node activation, node opening, evidence search, and pruning are written into the trace, together with state snapshots, search queries, activated pages, candidate indices, pruning reasons, and stop reasons. Beyond final accuracy, the trace can be used to analyze whether an error comes from missing initial pages, insufficient candidate generation, premature stopping, repeated search, wrong pruning, or the reader's failure to understand the final evidence. Subsequent ablations, efficiency statistics, and case studies are all based on this trace to explain behavioral differences across graph modes, controller configurations, and reader rendering strategies.

%% file: sections/experiments.tex
\section{Experiments}
\label{sec:experiments}

This section presents the experimental design, evaluation protocol, and result analysis for MAGE-RAG. In long-document multimodal QA, performance gains can easily be confounded with larger inputs. The experiments therefore examine task metrics, evidence scale, and query-time control behavior together. The central question is whether query-time evidence subgraph construction can organize evidence more effectively than fixed-granularity retrieval under the same reader and a controlled retrieval space.

\subsection{Experimental Setup}
\label{subsec:exp_setup}

\textbf{Research questions.}
The experiments are organized around four questions. \textbf{Q1:} How does MAGE-RAG perform on long-document multimodal QA compared with Direct MLLM, Text RAG, page-level visual RAG, and Graph/Agentic RAG? \textbf{Q2:} Are the gains of MAGE-RAG stable across single-page, cross-page, unanswerable questions, different evidence sources, and different answer formats? \textbf{Q3:} How do the initial page budget, controller iteration budget, and action budget affect performance and cost? \textbf{Q4:} What are the respective contributions of the multigranular evidence graph, online search, pruning, and structured rendering?

\textbf{Datasets.}
The main evaluation datasets are LongDocURL and MMLongBench-Doc. LongDocURL targets long-document understanding, numerical reasoning, and cross-element localization. It contains 2,325 QA samples and covers text, tables, images, layout, and cross-page evidence~\cite{longdocurl}. MMLongBench-Doc contains expert-annotated questions over long PDFs, with an emphasis on cross-page reasoning, multiple evidence sources, and unanswerable questions. It is an important benchmark for evaluating long-context document understanding~\cite{mmlongbenchdoc}. Together, the two datasets cover three representative scenarios: sparse evidence, cross-page evidence, and heterogeneous evidence modalities.

\textbf{Baselines.}
The compared methods are divided into four categories. The first is \emph{Direct MLLM}, which sends the pages allowed by the question directly to the reader without retrieval. The second is \emph{Text RAG}, including BM25~\cite{bm25} and ColBERTv2~\cite{colbertv2}, which rely on OCR text chunking and text retrieval. The third is \emph{Page-level Visual RAG}, including M3DocRAG~\cite{m3docrag}, VisRAG~\cite{visrag}, and EVisRAG/VisRAG 2.0~\cite{evisrag}, which retrieve page images. The fourth is \emph{Graph/Agentic RAG}, including MLDocRAG~\cite{mldocrag} and \(G^2\)-Reader~\cite{g2reader}, which explicitly introduce structured evidence organization or multi-step evidence exploration.

\textbf{Implementation protocol.}
All main comparisons use Qwen3-VL-8B-Instruct as the reader. This model belongs to the Qwen3-VL family, which supports visual documents, long contexts, and multi-image understanding~\cite{qwen3vl}. For RAG methods, the retrieval space is restricted to the documents or pages provided by the benchmark, without introducing pages outside the question scope. For LongDocURL, we report overall accuracy and fine-grained metrics over understanding, reasoning, locating, element type, and evidence pages. For MMLongBench-Doc, we report overall accuracy and overall F1, and further group results by single-page, cross-page, unanswerable, evidence source, document type, and answer format. Beyond final task metrics, the experiments also record reader input size, the number of evidence nodes, the number of images, and controller actions, so that the gains can be analyzed in terms of evidence organization rather than input expansion alone.

\input{tables/main_results_table}

\subsection{Main Results}
\label{subsec:main_results}

Table~\ref{tab:main_results} reports the main experimental results. The effectiveness of Text RAG mainly depends on OCR quality and whether chunk boundaries cover the answer evidence. In cases involving charts, images, layout localization, and cross-page evidence, such methods can easily lose visual structure. Page-level visual RAG preserves whole-page visual information and is more suitable for visually dense documents. Yet it still sends a fixed number of full-page images directly to the reader, making it difficult to distinguish high-value regions from distracting regions within a page. Graph/Agentic RAG begins to organize evidence explicitly, but existing methods often lean toward text graph reading, or lack budgeted state control that jointly aligns page-level visual evidence with local element evidence.

MAGE-RAG achieves 52.75 overall accuracy on LongDocURL, outperforming Direct MLLM, Text RAG, page-level visual RAG, and \(G^2\)-Reader. On MMLongBench-Doc, MAGE-RAG obtains 53.26 accuracy and 51.19 F1, exceeding \(G^2\)-Reader, which obtains 46.96 accuracy and 45.29 F1. Compared with MLDocRAG, MAGE-RAG achieves higher overall accuracy on MMLongBench-Doc and stronger performance on the cross-element group of LongDocURL. This indicates that query-time evidence subgraph construction not only compensates for evidence missed by fixed-granularity retrieval, but also organizes page-level and element-level evidence more effectively under the Graph/Agentic RAG setting.

Sample-level paired diagnostics are used to check whether the results are driven by a small number of outlier samples. On LongDocURL, the net-win counts of MAGE-RAG over BM25, ColBERTv2, EVisRAG, Direct MLLM, and M3DocRAG are 268, 373, 135, -97, and -102, respectively. Its gains over text retrieval and EVisRAG are more stable, whereas its advantage over full-page visual input and M3DocRAG is mainly reflected in aggregate metrics and hierarchical breakdowns. On MMLongBench-Doc, the net-win counts of MAGE-RAG over BM25, ColBERTv2, EVisRAG, and M3DocRAG are 194, 203, 70, and 117, respectively. The full-controller/full-graph configuration more consistently corrects the errors of fixed-granularity retrieval on this dataset.

\input{tables/main_breakdown_figure}

\textbf{Fine-grained breakdown.}
Fig.~\ref{fig:main_breakdown} reports the hierarchical results for Q2. On LongDocURL, MAGE-RAG achieves the best results on the understanding, reasoning, and single-page groups; its multi-page result is close to Direct MLLM but does not exceed it. The locating group is close to M3DocRAG, suggesting that localization questions still depend strongly on page-level visual grounding. On MMLongBench-Doc, MAGE-RAG achieves the best results on both single-page and cross-page groups. Its cross-page accuracy is 34.70, higher than 24.87 for M3DocRAG and 24.51 for EVisRAG. The unanswerable group requires separate interpretation: BM25 and ColBERTv2 score higher on this group, but their overall accuracy, F1, and single/cross-page accuracy are clearly lower. A high unanswerable score alone therefore does not indicate stronger overall document understanding. Overall, the gains of MAGE-RAG mainly come from task understanding, reasoning, and cross-page evidence organization on answerable samples.

\subsection{Budget and Ablation Analysis}
\label{subsec:budget_ablation}

\textbf{Budget sensitivity.}
The budget sensitivity experiments vary three axes: the number of initial pages \(k\), the number of controller iterations \(T\), and the maximum number of actions per iteration \(a\). Here, \(T=0\) means that only initial page grounding and deterministic rendering are performed, without running the online controller. This corresponds to a page-level entry followed by fixed evidence rendering. As \(T\) and \(a\) increase, the system can open more nodes, issue focused search requests, or prune low-value nodes. The analysis tracks both performance curves and logical cost curves, in order to distinguish effective evidence selection from simply enlarging the context.

\input{tables/mmlongbench_magerag_budget_table}

\begin{figure}[t]
\centering
\includegraphics[width=\columnwidth]{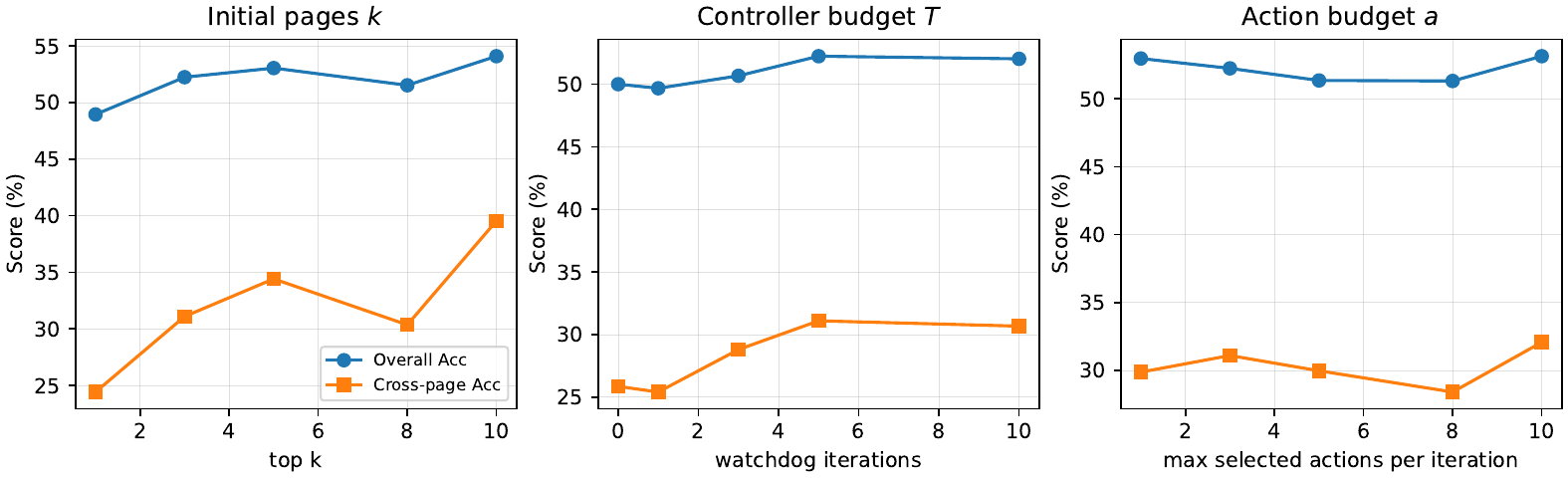}
\caption{Budget sensitivity analysis of MAGE-RAG on MMLongBench-Doc. The three groups of curves vary the number of initial pages $k$, the number of controller rounds $T$, and the number of actions per round $a$, while reporting both overall accuracy and cross-page accuracy.}
\label{fig:magerag_budget_curves}
\end{figure}

Table~\ref{tab:magerag_budget} and Fig.~\ref{fig:magerag_budget_curves} show the budget sweep on MMLongBench-Doc. In general, the benefit of increasing the budget is non-monotonic. Under $T=5,a=3$, $k=5$ reaches 53.05 overall accuracy and 34.42 cross-page accuracy, outperforming $k=1$ and $k=3$; however, the cross-page accuracy of $k=8$ drops to 30.36. A larger $k=10$ further improves the cross-page metric, but reduces the unanswerable metric. The controller iteration budget shows a saturation pattern: with $k=3,a=3$, increasing \(T\) from 0 to 5 improves cross-page accuracy from 25.86 to 31.10, while \(T=10\) brings no further gain. The action budget also has an upper-limit effect. A small number of actions can already produce competitive results, whereas an overly large per-round action budget may introduce more candidate evidence and decision noise. The budget primarily changes the search trajectory of the evidence state; simply expanding the context does not yield stable gains.

\textbf{Ablation variants.}
The ablation study includes two representative groups of comparisons. The first group studies progressive capability: \emph{Top-k Page Only}, \emph{Top-k Page with Node Rendering}, \emph{Graph Neighbor Expansion}, and \emph{Dynamic Controller without Search}. This group examines how the method moves from a page entry toward graph-driven evidence construction, and checks the effect of online search actions on cross-page questions. The second group studies graph structure: under the full controller, the retrieval space is restricted to \emph{Page Only}, \emph{Containment Only}, or \emph{Full Graph}, respectively, to analyze how different evidence graph structures affect final performance.

\input{tables/mmlongbench_magerag_ablation_table}

Table~\ref{tab:magerag_ablation} summarizes the representative ablation results. The improvement from top-k page only to top-k page with node rendering indicates that local node rendering can provide a more compact evidence representation for the page-level entry. Graph neighbor expansion achieves the highest overall, F1, cross-page, and unanswerable metrics in this batch of results, showing that expanding adjacent evidence along the evidence graph already covers a substantial portion of question needs. Dynamic controller without search has a lower cross-page metric, which suggests that online search remains important for completing dispersed evidence. In the graph-structure ablation, both page-only and containment-only settings retain part of the gains. These results are more consistent with the interpretation that graph structure and control strategy jointly affect evidence organization, rather than with an explanation based on a single edge type.

\subsection{Trace-based and Efficiency Analysis}
\label{subsec:trace_efficiency}

\textbf{Trace-based analysis.}
The query process of MAGE-RAG produces an auditable evidence-control trajectory. Trace-based analysis counts activated pages, opened nodes, search count, prune count, and the number of iterations, and further splits the statistics by correct and incorrect samples. For incorrect samples, these trajectories provide inspectable attribution clues: whether the initial retrieval hits the answer page, whether focused search recovers missing evidence, whether the controller stops too early or prunes incorrectly, and whether the final reader input already contains answer-supporting evidence.

\input{tables/mmlongbench_magerag_trace_stats_table}

Table~\ref{tab:magerag_trace_stats} and Fig.~\ref{fig:magerag_trace_stats} report trace statistics from a representative run. Correct samples open 17.57 nodes on average, activate 5.28 pages, trigger 1.53 search actions, and send 21.54 images to the reader. The corresponding values for incorrect samples are 16.58, 5.11, 1.41, and 20.42. The differences between the two groups are small but directionally consistent. Incorrect samples have slightly larger average iteration counts and prune counts, indicating that a longer or more complex trace does not by itself guarantee success. The source of an error still needs to be judged together with the concrete stopping reason, answer-page hit status, and final reader input.

\textbf{Efficiency analysis.}
This work does not use wall-clock time as the main efficiency conclusion, because model serving concurrency, queue length, and backend load can substantially affect latency. Table~\ref{tab:magerag_efficiency} reports logical cost from the same representative run used for the trace statistics. In this run, each sample sends 20.99 images, 4.05 context pages, 21.05 context nodes, and 49.02 content parts to the reader on average, and triggers 3.23 LLM calls per sample. The aggregate efficiency is 0.025 score/input image and 0.162 score/LLM call. Correct samples have slightly wider reader inputs, with 21.54 images, 4.11 pages, and 21.59 nodes on average; incorrect samples have 20.42 images, 3.99 pages, and 20.50 nodes. This difference is consistent with the trace statistics in Table~\ref{tab:magerag_trace_stats}: correct samples do not rely on substantially larger final inputs, and the results should be interpreted together with the budget and ablation analyses.

\input{tables/mmlongbench_magerag_efficiency_table}

\textbf{Theoretical view.}
From the perspective of experimental interpretation, the theoretical assumption behind MAGE-RAG can be summarized as budgeted subgraph selection: given an initial page set and an evidence graph, the method attempts to increase answerability, reduce noise, and control reader cost within a limited action budget. The budget curves, ablations, and trace analysis correspond to three observable aspects of this objective. The budget curves examine the cost-aware trade-off, the ablations test the subgraph search space and action set, and the trace analysis checks whether the online selection process is consistent with evidence coverage and error types. The experimental section therefore interprets Eq.~\eqref{eq:objective} only through measurable metrics, without introducing an additional assumption of end-to-end optimality.

%% file: tables/main_results_table.tex
\begin{table*}[t]
\centering
\caption{Main experimental results on LongDocURL and MMLongBench-Doc. TXT, LAY, CHA, FIG, and TAB denote text, layout, chart, figure, and table evidence sources. SP/SIN, MP/MUL, CE, and UNA denote single-page, multi-page, cross-element, and unanswerable subsets. Values are reported as percentages.}
\label{tab:main_results}
\setlength{\tabcolsep}{2.7pt}
\renewcommand{\arraystretch}{1.05}
\scriptsize
\resizebox{\textwidth}{!}{%
\begin{tabular}{ll|cccccccc|cccccccccc}
\hline\hline
\multicolumn{2}{c|}{Method} &
\multicolumn{8}{c|}{LongDocURL} &
\multicolumn{10}{c}{MMLongBench-Doc} \\
\cline{1-20}
Method & Type &
\multicolumn{4}{c|}{Evidence Source} &
\multicolumn{3}{c|}{Evidence Page} & Overall &
\multicolumn{5}{c|}{Evidence Source} &
\multicolumn{3}{c|}{Evidence Page} & Acc & F1 \\
\cline{3-9}\cline{11-18}
 & & TXT & LAY & FIG & TAB & SP & MP & CE & Acc
 & TXT & LAY & CHA & TAB & FIG & SIN & MUL & UNA & & \\
\hline
Direct MLLM & Direct MLLM & 57.49 & 41.74 & 50.21 & 47.29 & 51.91 & \underline{48.44} & 48.32 & 50.07 & \textbf{48.98} & \textbf{50.34} & 38.01 & \underline{53.67} & \underline{41.39} & \underline{54.83} & \textbf{38.58} & 27.62 & 43.05 & 43.63 \\
\hline
BM25 & Text RAG & 44.74 & 32.54 & 37.68 & 30.71 & 38.24 & 34.57 & 37.84 & 36.26 & 20.84 & 14.73 & 12.36 & 22.38 & 3.75 & 21.74 & 8.82 & \underline{85.66} & 31.16 & 22.03 \\
ColBERTv2 & Text RAG & 41.81 & 29.71 & 32.46 & 22.67 & 30.08 & 33.27 & 32.52 & 31.74 & 21.26 & 15.83 & 11.57 & 16.04 & 4.10 & 18.70 & 9.70 & \textbf{87.30} & 30.56 & 20.43 \\
\hline
M3DocRAG & Page-level Visual RAG & 56.20 & \underline{43.04} & \underline{50.49} & 45.62 & 53.47 & 45.77 & \underline{48.51} & 49.35 & 35.48 & 35.62 & 37.32 & 38.45 & 39.25 & 50.40 & 24.87 & 34.29 & 38.21 & 37.52 \\
EVisRAG & Page-level Visual RAG & 48.56 & 34.21 & 41.10 & 41.41 & 47.36 & 37.01 & 42.09 & 41.84 & 37.25 & 32.91 & 41.83 & 41.54 & 35.23 & 52.03 & 24.51 & 50.00 & 42.33 & 40.65 \\
\hline
G2Reader & Graph/Agentic RAG & 57.82 & 38.38 & 46.69 & \underline{50.95} & 54.31 & 47.44 & 45.24 & 50.67 & 40.99 & 32.61 & 41.48 & \textbf{57.30} & 31.25 & 53.91 & 30.69 & 58.87 & 46.96 & \underline{45.29} \\
MLDocRAG & Graph/Agentic RAG & \textbf{65.10} & 39.40 & 48.30 & 41.10 & \textbf{66.90} & \textbf{56.30} & 23.40 & \underline{50.80} & \underline{47.20} & 37.80 & \underline{42.70} & 41.30 & 31.90 & 52.60 & 26.40 & 71.50 & \underline{47.90} & -- \\
MAGE-RAG & Graph/Agentic RAG & \underline{59.82} & \textbf{45.08} & \textbf{51.00} & \textbf{51.60} & \underline{58.44} & 47.69 & \textbf{51.95} & \textbf{52.75} & 46.02 & \underline{46.70} & \textbf{45.41} & 51.17 & \textbf{42.46} & \textbf{60.22} & \underline{34.70} & 68.03 & \textbf{53.26} & \textbf{51.19} \\
\hline\hline
\end{tabular}%
}
\vspace{1mm}
\end{table*}

%% file: tables/main_breakdown_figure.tex
\begin{figure*}[t]
\centering
\includegraphics[width=\textwidth]{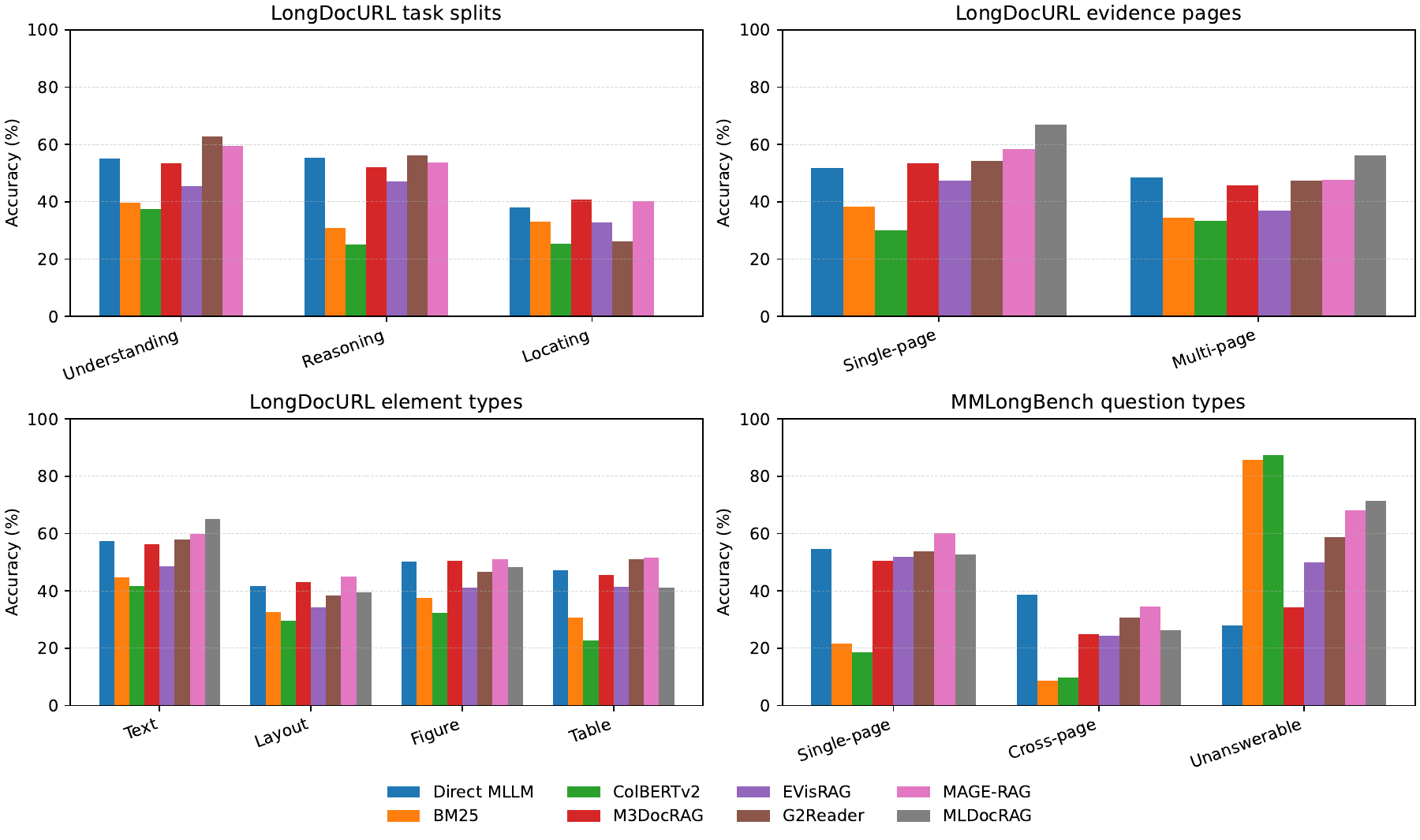}
\caption{Fine-grained breakdown results on LongDocURL and MMLongBench-Doc. LongDocURL is split by task type, number of evidence pages, and element type; MMLongBench-Doc is split by single-page, cross-page, and unanswerable questions.}
\label{fig:main_breakdown}
\end{figure*}

%% file: tables/mmlongbench_magerag_budget_table.tex
\begin{table}[t]
\centering
\caption{Budget-sensitivity results of MAGE-RAG on MMLongBench-Doc. All runs use the full controller and full evidence graph; Acc/F1 are reported as percentages.}
\label{tab:magerag_budget}
\scriptsize
\resizebox{\columnwidth}{!}{%
\begin{tabular}{cccccccc}
\hline
$k$ & $T$ & $a$ & Acc & F1 & Single & Cross & Unans. \\
\hline
1 & 5 & 3 & 48.95 & 45.75 & 55.87 & 24.42 & 73.77 \\
3 & 0 & 3 & 50.01 & 46.82 & 55.30 & 25.86 & 76.64 \\
3 & 0 & 5 & 49.77 & 46.98 & 56.52 & 24.35 & 75.82 \\
3 & 1 & 3 & 49.68 & 46.75 & 56.76 & 25.41 & 73.36 \\
3 & 1 & 5 & 48.85 & 47.04 & 56.83 & 26.09 & 68.03 \\
3 & 3 & 3 & 50.68 & 48.16 & 58.27 & 28.80 & 69.67 \\
3 & 3 & 5 & 51.29 & 49.08 & 60.82 & 26.83 & 69.90 \\
3 & 5 & 1 & 52.97 & 49.80 & 61.79 & 29.87 & 71.72 \\
3 & 5 & 3 & 52.25 & 49.44 & 59.43 & 31.10 & 70.90 \\
3 & 5 & 5 & 51.35 & 48.94 & 59.32 & 29.97 & 69.26 \\
3 & 5 & 8 & 51.32 & 48.34 & 58.77 & 28.41 & 72.13 \\
3 & 5 & 10 & 53.14 & 50.02 & 59.55 & 32.07 & 72.93 \\
3 & 10 & 3 & 52.04 & 49.45 & 59.11 & 30.67 & 71.72 \\
5 & 5 & 3 & 53.05 & 51.12 & 60.16 & 34.42 & 67.62 \\
5 & 10 & 5 & 53.26 & 51.19 & 60.22 & 34.70 & 68.03 \\
8 & 5 & 3 & 51.53 & 49.34 & 61.41 & 30.36 & 65.16 \\
10 & 5 & 3 & 54.10 & 52.61 & 59.97 & 39.55 & 64.29 \\
\hline
\end{tabular}%
}
\end{table}

%% file: tables/mmlongbench_magerag_ablation_table.tex
\begin{table}[t]
\centering
\caption{Capability and graph-structure ablations of MAGE-RAG on MMLongBench-Doc. All runs fix $k=3,T=10,a=5$; Acc/F1 are reported as percentages.}
\label{tab:magerag_ablation}
\scriptsize
\resizebox{\columnwidth}{!}{%
\begin{tabular}{llccccc}
\hline
Controller & Graph & Acc & F1 & Single & Cross & Unans. \\
\hline
Topk Page Only & Full Graph & 48.75 & 45.71 & 55.30 & 24.31 & 73.36 \\
Topk Page With Node Rendering & Full Graph & 49.71 & 47.38 & 57.09 & 26.42 & 71.31 \\
Graph Neighbor Expansion & Full Graph & 53.88 & 50.20 & 59.83 & 32.11 & 76.23 \\
Dynamic Controller No Search & Full Graph & 47.95 & 45.52 & 55.76 & 22.15 & 72.54 \\
Full & Containment Only & 51.21 & 48.17 & 58.42 & 28.27 & 71.72 \\
Full & Layout Graph & 50.71 & 47.98 & 58.08 & 29.11 & 68.85 \\
Full & Page Only & 52.36 & 49.35 & 58.62 & 31.70 & 72.13 \\
Full & Semantic Graph & 53.36 & 49.91 & 60.37 & 29.19 & 75.23 \\
Full & Structural Graph & 50.05 & 47.51 & 58.69 & 26.53 & 69.01 \\
Dynamic Controller No Prune & Full Graph & 52.99 & 50.26 & 60.64 & 31.45 & 70.48 \\
\hline
\end{tabular}%
}
\end{table}

%% file: tables/mmlongbench_magerag_trace_stats_table.tex
\begin{table}[t]
\centering
\caption{Trace statistics of MAGE-RAG on MMLongBench-Doc. Statistics are computed from a representative run with $k=3,T=5,a=3$; averages are split by answer-level correct and incorrect samples.}
\label{tab:magerag_trace_stats}
\scriptsize
\resizebox{\columnwidth}{!}{%
\begin{tabular}{lcccccccc}
\hline
Group & Count & Steps & Iter. & Pages & Open & Prune & Search & Images \\
\hline
Correct & 549 & 64.06 & 3.13 & 5.28 & 17.57 & 1.64 & 1.53 & 21.54 \\
Incorrect & 542 & 61.20 & 3.34 & 5.11 & 16.58 & 1.73 & 1.41 & 20.42 \\
\hline
\end{tabular}%
}
\end{table}

%% file: tables/mmlongbench_magerag_efficiency_table.tex
\begin{table}[t]
\centering
\caption{Logical-cost statistics of MAGE-RAG on MMLongBench-Doc. Statistics are computed from a representative run with $k=3,T=5,a=3$; Score/Image and Score/LLM divide the aggregate score by the corresponding aggregate cost.}
\label{tab:magerag_efficiency}
\scriptsize
\resizebox{\columnwidth}{!}{%
\begin{tabular}{lccccccccc}
\hline
Group & Count & Acc & Images & Pages & LLM & Parts & Nodes & Score/Image & Score/LLM \\
\hline
Overall & 1091 & 52.25 & 20.99 & 4.05 & 3.23 & 49.02 & 21.05 & 0.025 & 0.162 \\
Correct & 549 & 100.00 & 21.54 & 4.11 & 3.13 & 50.19 & 21.59 & 0.046 & 0.320 \\
Incorrect & 542 & 3.87 & 20.42 & 3.99 & 3.34 & 47.83 & 20.50 & 0.002 & 0.012 \\
\hline
\end{tabular}%
}
\end{table}

%% file: sections/conclusion.tex
\section{Conclusion}
\label{sec:conclusion}

This paper formulates evidence organization in long-document multimodal QA as query-time evidence subgraph construction under budget constraints. MAGE-RAG combines a page-level visual entry point, a multigranular evidence graph, an online evidence controller, and structured reader rendering, so that page retrieval, local evidence expansion, and reader input construction become connected but distinct steps for organizing sparse, cross-page, and visually heterogeneous evidence in long PDFs.

On LongDocURL and MMLongBench-Doc, the main results, budget sweeps, component ablations, and trace statistics show that the gains of MAGE-RAG do not simply come from enlarging the context. Starting from page-level entries, the system opens local nodes according to the question, completes evidence along graph relations, and removes low-value content before final rendering. Its trace records page activation, node opening, focused search, pruning, and stopping reasons for error attribution and debugging.

The upper bound of this framework still depends on the quality of the offline evidence graph, and the controller introduces additional evaluator calls and state-maintenance cost. MAGE-RAG is better suited to long-document questions with sparse evidence, cross-page dependencies, or complex visual layouts, and may not be cost-effective for all short-document or single-hop cases. Future work should improve graph construction for tables, charts, captions, and cross-page section structures, reduce repeated searches and ineffective openings, and extend trace-based diagnosis into systematic error attribution.

%% file: sections/acknowledgment.tex
\section*{Acknowledgment}

The authors thank their advisors for guidance on the research topic, method design, and paper writing, and thank lab members for help with code reproduction, experiments, result checking, and technical discussions. Their suggestions helped clarify the motivation, experimental setup, and method presentation of this work.

\textbf{AI-generated content acknowledgment.}
Generative AI tools, including ChatGPT and OpenAI Codex, were used mainly for code writing and debugging, experimental-design discussion, and literature collection. All AI-generated content was manually reviewed, revised, and confirmed by the authors, who remain responsible for the research design, interpretation of results, arguments, and final conclusions.